\documentclass{aa}

\usepackage{amsmath}
\usepackage{flexisym}
\usepackage{graphicx}

\usepackage{booktabs}
\usepackage{tabularx}
\usepackage[breaklinks]{hyperref}
\usepackage[table]{xcolor}
\usepackage{multirow}
\usepackage{lipsum}
\usepackage{tikz}
\usetikzlibrary{shapes.geometric, arrows}

\def \Lsixum {{{\rm L}_{6\mu \rm  m}}}
\def \Llofar {$\rm L_{\rm 150MHz}$}

\def \ergs {{$\rm ergs \, s^{-1}$}}

\defcitealias{CR21}{CR21}
\usepackage{graphicx}
\usepackage{txfonts}
\begin{document} 


\title{Ubiquitous radio emission in quasars: predominant AGN origin and a connection to jets, dust and winds}
\titlerunning{Ubiquitous radio emission in quasars}


\author{G. Calistro Rivera \inst{1} \and D. M. Alexander\inst{2} \and  C. M. Harrison \inst{3}  \and V. A. Fawcett  \inst{3} \and P. N. Best \inst{4} \and W.L. Williams \inst{8} \and M. J. Hardcastle\inst{5} \and  D. J. Rosario \inst{3}  \and D. J. B. Smith \inst{5}  \and M. I. Arnaudova \inst{5} \and E. Escott  \inst{2} \and  G. G\"urkan\inst{5,9}  \and R. Kondapally \inst{4}  \and G. Miley  \inst{6} \and L.~K. Morabito \inst{2,10} \and J. Petley  \inst{2} \and I. Prandoni \inst{7} \and H.J.A. R\"ottgering  \inst{6} \and B.-H. Yue  \inst{4,6}  
}%

\institute{European Southern Observatory (ESO), Karl-Schwarzschild-Stra\ss e 2, 85748 Garching bei M\"unchen, Germany \\\email{gcalistrorivera@gmail.com} 
\and
Centre for Extragalactic Astronomy, Department of Physics, Durham University, Durham DH1 3LE, UK
\and
School of Mathematics, Statistics and Physics, Newcastle University, Newcastle upon Tyne, NE1 7RU, UK
\and
Institute for Astronomy, University of Edinburgh, Royal Observatory, Blackford Hill, Edinburgh EH9 3HJ, UK
\and
Centre for Astrophysics Research, University of Hertfordshire, College Lane, Hatfield AL10 9AB, UK
\and
Leiden Observatory, Leiden University, PO Box 9513, 2300 RA Leiden, The Netherlands
\and
INAF - Istituto di Radioastronomia, Via P. Gobetti 101, 40129 Bologna (Italy)
\and
SKA Observatory, Jodrell Bank, Lower Withington, Macclesfield, SK11 9FT, UK
\and
CSIRO Space and Astronomy, ATNF, PO Box 1130, Bentley, WA 6102, Australia
\and 
Institute for Computational Cosmology, Department of Physics, University of Durham, South Road, Durham DH1 3LE, UK
}

  \abstract
  {  
We present a comprehensive study of the physical origin of radio emission in optical quasars at redshifts  $z<2.5$. We focus particularly on the associations between compact radio emission, dust reddening, and outflows identified in our earlier work. Leveraging the deepest low-frequency radio data available to date (LoTSS Deep DR1), we achieve radio detection fractions of up to 94\%, demonstrating the virtual ubiquity of radio emission in quasars, and a continuous distribution in radio loudness.
Through our analysis of radio properties, combined with spectral energy distribution modelling of deep multiwavelength photometry, we establish that the primary source of radio emission in quasars is the AGN, rather than star formation. 
Modelling the dust reddening of the accretion disk emission shows a continuous increase in radio detection in quasars as a function of the reddening parameter E(B-V), suggesting a causal link between radio emission and dust reddening.
Confirming previous findings, we observe that the radio excess in red quasars is most pronounced for sources with compact radio morphologies and intermediate radio loudness. 
We find a significant increase in [\textsc{Oiii}] and \textsc{Civ} outflow velocities for red quasars not seen in our control sample, with particularly powerful [\textsc{Oiii}] winds in those around the radio-quiet/radio-loud threshold. 
Based on the combined characterisation of radio, reddening, and outflow properties in our sample, we favour a model in which the compact radio emission observed in quasars originates in compact radio jets and their interaction with a dusty, circumnuclear environment. 
In particular, our results align with the theory that jet-induced winds and shocks resulting from this interaction are the origin of the enhanced radio emission in red quasars. Further investigation of this model is crucial for advancing our understanding of quasar feedback mechanisms and their role in galaxy evolution.

  }
   \keywords{}

   \maketitle

\section{Introduction}

Quasars (QSOs) were first discovered as bright radio sources with luminous star-like compact optical counterparts. Through optical follow-up spectroscopy \citep{schmidt63} they were then identified to be extragalactic objects, powered by accreting supermassive black holes (SMBHs). 
Despite their discovery in association with powerful radio emission, later found to be predominantly due to radio jets and lobes, it would  soon become clear that the majority of QSOs actually lacked radio detections at the sensitivities of the telescopes at the time. 
With new-generation radio interferometers of high sensitivity and large sky coverage, such as the Low-Frequency Array \citep[LOFAR; ][]{vanhaarlem+13}, we are now able to investigate the radio properties of the QSO population to the lowest radio luminosities \citep[e.g., ][Arnaudova et al. sub, Yue et al. sub]{gurkan19, macfarlane21}.

Despite these advancements, the origin of the faint radio emission in QSOs  remains strongly debated in the community \citep{bonzini15, padovani16,  panessa19, chen23}.
The question on the nature of the radio emission in QSOs is challenging since complementary multiwavelength and multi-scale information is required to disentangle the primary physical origin amid various processes that can give rise to radio emission.
On the one hand, synchrotron emission can originate in star forming regions through the interaction of the high-energy electron plasma from supernova events and magnetic fields, and therefore can be directly linked to massive star formation at galactic scales \citep[e.g., ][]{helou85, bell03}.  
On the other hand, the radio emission in QSOs can be of AGN origin, and connected to the black hole accretion responsible for the energetic optical radiation that defines QSOs. 
Here, physical mechanisms which could produce the synchrotron radio emission are low-power, small-scale versions of the kpc-scale jets observed in traditionally `radio-loud' quasars \citep[e.g. ][]{jarvis19, hartley19}, shocked electrons produced by the interactions of the interstellar medium (ISM) with jets and/or powerful winds \citep[e.g. ][]{zakamska14, nims15, hwang18}, as well as hot electrons from the X-ray-emitting corona \citep[e.g. ][]{laor08, baldi21, chen23}.
The identification of the main mechanism(s) responsible for the radio emission is key to characterise the mechanical energy and momentum released as a function of the BHs properties.
This is in turn essential to constrain the overall impact of accreting SMBHs on their host galaxies (i.e., black hole/AGN feedback) and is therefore a potential avenue to observationally characterise feedback mechanisms in galaxy evolution.

Traditionally, obscuration in AGN has been ascribed to the dusty structure around the accretion disk known as the torus, and to its orientation with respect to the line of sight \citep[][]{antonucci93}.
Recently, this picture is evolving, as increasing evidence suggests that the evolutionary stage of the AGN is also an important factor for obscuration  \citep[e.g.][]{hopkins04, alexander12, ricci17}.
In particular, fundamental differences have been seen in the radio emission between optical reddened QSOs and the bulk of the QSO population with average blue colours \citep[][]{klindt19, fawcett20,fawcett21, fawcett23, rosario20, rosario21, glikman22}.
This finding appears in tension with models that ascribe obscuration exclusively to orientation of the dusty torus. 
Instead, it may be consistent with the notion that some  QSOs are obscured by polar dust, potentially distributed within winds \citep[e.g.,][]{stalevski19, honig17} and/or galaxy-scale material linked to different evolutionary phases \citep[e.g.,][]{circosta19, andonie23}.
Indeed, recent studies suggest that there is a close connection between QSO reddening and increased level of outflows \citep{zakamska16, hamann17,  bischetti17, dipompeo18, mehdipour18, perrotta19, temple19, villar20, rojas20, CR21, stacey22},
where the dust responsible for the red colour in QSOs could potentially reside in circumnuclear dusty winds \citep{CR21}.
Similarly, studies have also focused on the connection between the radio emission and outflow properties in QSOs \citep[e.g, ][]{jarvis21, girdhar22, baldi21}, finding a positive correlation \citep[e.g., ][Escott et al., in prep.; Petley
et al., in prep]{mullaney13, wang22}, in particular for compact and GHz-peaked, potentially younger, radio sources \citep{molyneux19, kukreti23}.

While  most of these  studies have individually explored the connection between radio emission and either reddening or outflows, drawing comprehensive conclusions has proven challenging due to the heterogeneity of data sets, and sometimes limited sample sizes and multiwavelength coverage.
In this paper we leverage the deepest radio data to date from the LOFAR telescope \citep[LoTSS Deep DR1; ][]{tasse21, sabater21}, in combination with a comprehensive multiwavelength characterisation, to investigate the physical origin of the radio emission in QSOs.
Crucially, using a single consistent sample, we investigate the connection between the radio, reddening and outflow properties of QSOs.
In particular, we build upon a detailed analysis of the multiwavelength spectral energy distributions (SEDs) and optical spectral properties of QSOs presented by \citet{CR21}; here after \citetalias{CR21}.
Throughout this work, we adopt a cosmology with $H_{0}=70$km s$^{-1}$ Mpc$^{-1}$, $\Omega_{\rm m}=0.3$ and $\Omega_{\Lambda}=0.7$.


\begin{figure}
\centering
\centering
    \includegraphics[trim={0.2cm 0.2cm 0.2cm 0.2cm},clip, width=\linewidth]{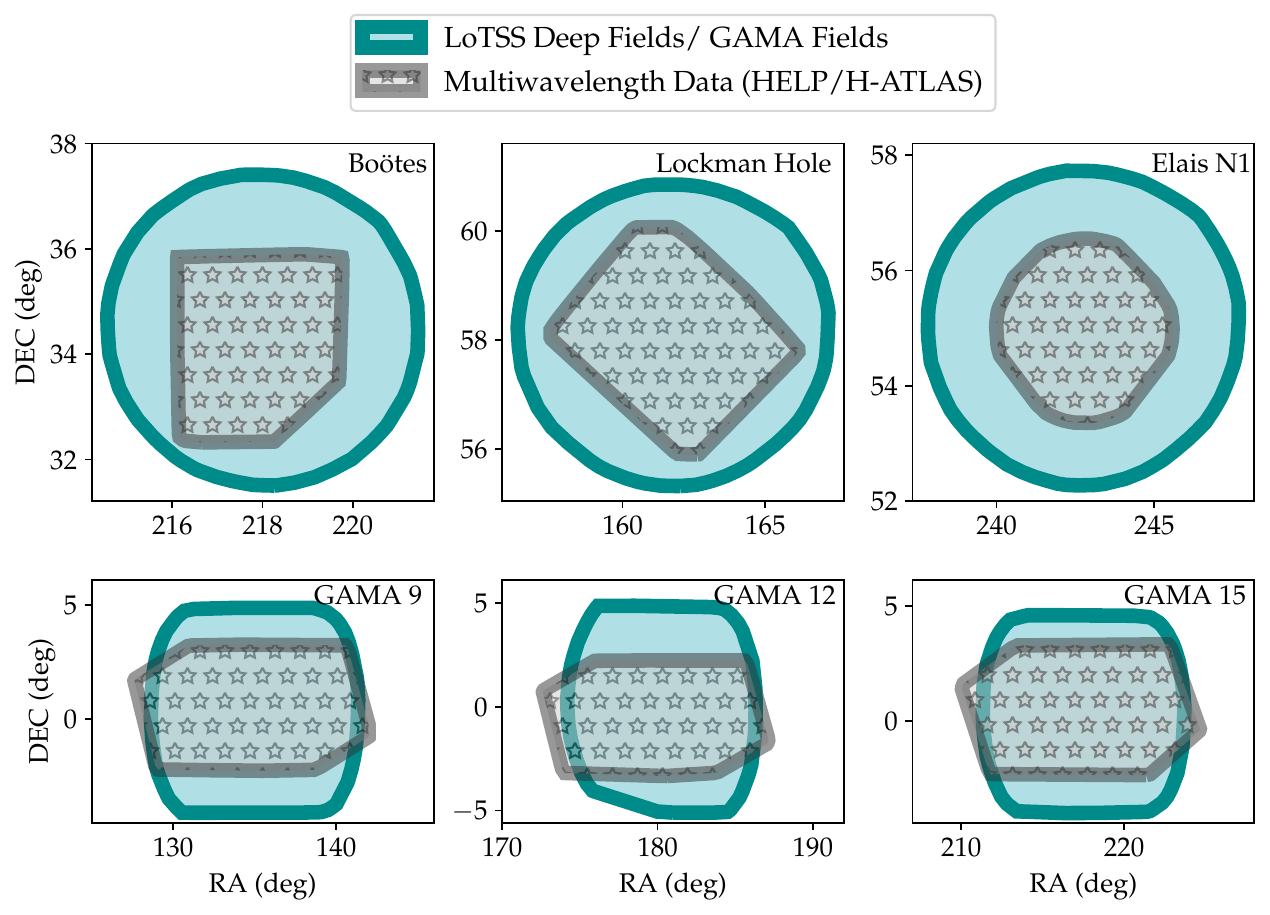}
    \includegraphics[trim={ 0.2cm 1cm 2.3cm 2.1cm},clip,width=\linewidth]{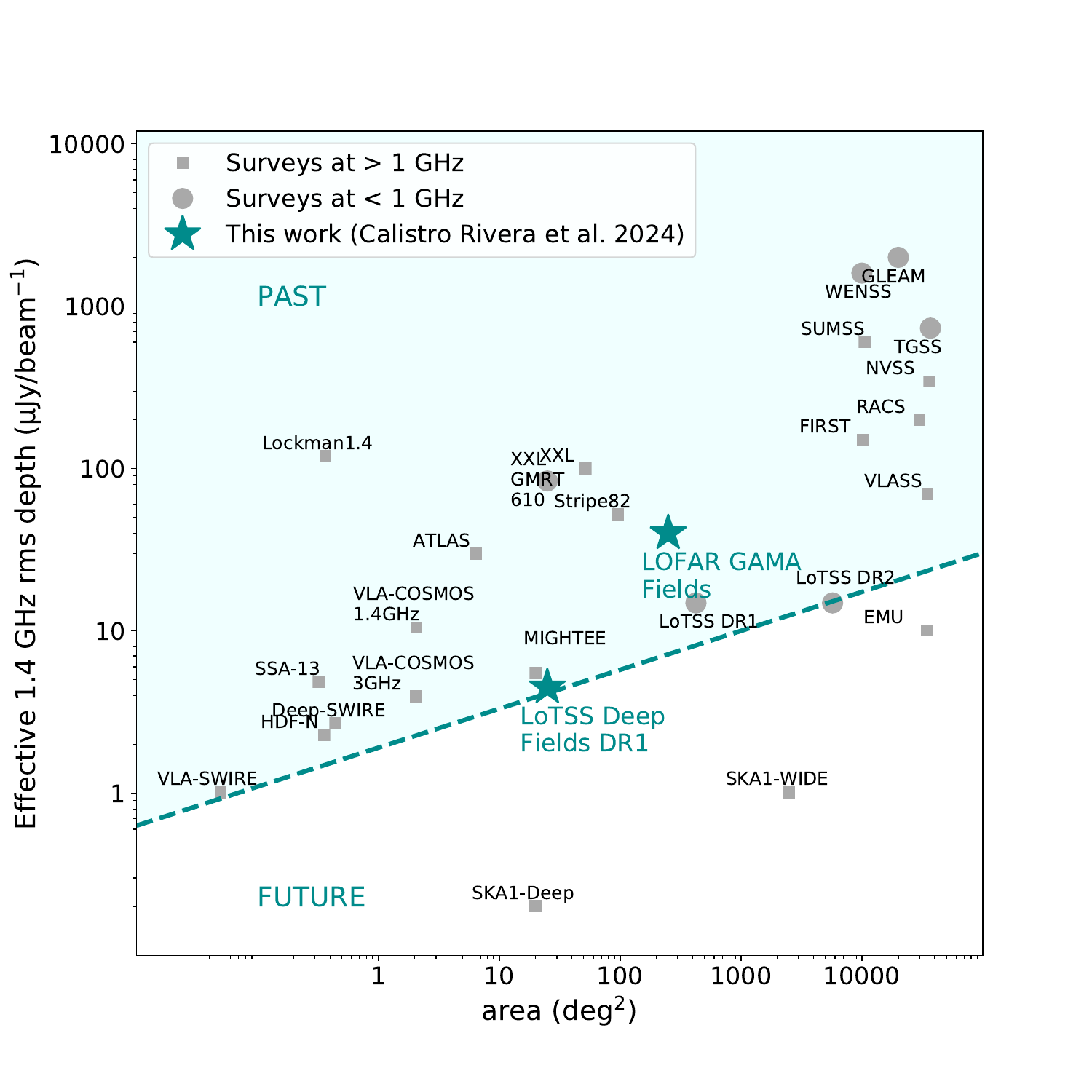}
    \caption{ Radio and multiwavelength data used in this paper. The upper panel shows the sky coverage of the LOFAR pointings (cyan area) in the 3 LoTSS Deep Fields DR1 (upper sub-panel) and the GAMA fields (lower sub-panel). The sky areas with deep multiwavelength coverage are overplotted (grey stars). The lower panel show the rms sensitivities and sky coverage of the LOFAR data used in this work, i.e. the LoTSS Deep Fields and the LOFAR GAMA Fields, compared to other past and future radio surveys. To enable the comparison, all survey depths are converted to a 1.4 GHz equivalent rms assuming a spectral index of $\alpha = 0.7$. }
    \label{fig:fields}
\end{figure}

\section{Observations and analysis}

\subsection{Sample selection, multiwavelength data, and SED fiting}\label{sec:selection}

The selection process for the QSO sample used in this investigation is described in detail by \citetalias{CR21} as well as in our previous SDSS QSO studies \citep{klindt19, rosario20, fawcett20, fawcett22}. 
In summary, we select our sample from the SDSS DR14 Quasar catalogue \citep{paris18}, which has detailed photometric and emission line measurements \citep{rakshit20}. 
Adopting the rest-frame luminosity at 6 $\mu$m ($\Lsixum$) as the proxy for QSO intrinsic luminosities which is most adequate for our study of optical reddening, we further select only sources with  WISE \citep[(W1, W2 $\&$ W3)][]{wright10} 
detections at SNR$>3$.
Then, based on SDSS $ugriz$ bands, we use the redshift-dependent $g^*-i^*$ colours to define the populations of red QSOs (10$\%$ reddest) and a control sample of `normal' QSOs (50$\%$ around the distribution median).
As demonstrated in our previous studies, this selection ensures a robust and unbiased comparison between red QSOs and a consistently selected control sample. 
Finally, the most significant cut made by \citetalias{CR21} on the SDSS QSO parent sample was the sky coverage.
As illustrated in Figure \ref{fig:fields}, six survey fields were chosen to achieve the combined availability of deep radio observations and multiwavelength data across all frequencies from the radio to X-rays for all QSOs.
They comprise a total area of 175 deg$^2$, including three surveys in the northern sky which are part of the LoTSS Deep Field surveys: the Bo\"otes field, the Lockman-Hole Field, and the ELAIS-N1 Fields \citep{tasse21, sabater21}, and three areas in the equatorial sky covered by the GAMA fields \citep{driver11}, as shown in the upper panel of Figure \ref{fig:fields}. 
In the overall LOFAR fields shown by the light cyan areas, we have 4436 QSOs at $z<2.5$, from which 755 are red QSOs and 3681 represent the control sample (ALLQSOs sample, see Table \ref{tab:samples}). 

To obtain a detailed multiwavelength characterisation of the QSOs in this study, the SDSS $ugriz$ and WISE photometric bands used in the selection, were complemented with abundant FIR-to-UV aperture-matched photometry compiled by \citet{kondapally20} for the LoTSS Deep Fields, and the \textit{Herschel} Extragalactic Legacy Project HELP project \citep{shirley19} for the equatorial GAMA fields.
A more detailed description of the multiwavelength photometry consisting of 25--30 photometric data points, from the FIR to the UV (including SDSS and WISE photometry), and the Bayesian SED modelling were presented by \citetalias{CR21}.
The fraction of the fields (LoTSS Deep + GAMA) with multiwavelength coverage (MWQSOs sample) include 1789 QSOs, from which 306 are red QSOs and 1483 represent the control sample. 
To disentangle the different physical components, and infer the relevant physical parameters we used the Bayesian \textsc{AGNfitter} code \citep{CR16}, which applies an MCMC approach to sample the posterior distributions of all physical parameters. 
The physical model for the host galaxy component in \textsc{AGNfitter} consists of the stellar populations in the NIR/optical/UV \citep[models by][]{bruzual03} and the cold dust emission in the IR/sub-mm \citep[models by ][]{schreiber18}.
The galaxy infrared luminosities used as tracers of the star formation rates (SFR) were estimated by integrating the cold-dust model within 8-1000 $\mu$m.
The AGN emission was estimated by the addition of the accretion disk emission in the optical/UV \citep[][]{richards06} and the hot dust/torus emission \citep[models by][]{silva04}.
In particular, the dust-reddening of the accretion disk (also called the Big Blue Bump, BBB) is modelled following a Small Magellanic Cloud dust attenuation law \citep{prevot84}, and parametrized by the reddening parameter E(B-V)$_{\rm BBB}$.

\begin{table}[]
    \centering
    \begin{tabular}{c|c|c|c}
         Samples &    LOFAR    & SDSS    & LOFAR \\
                 &    Fields   &  QSOs   & 3$\sigma$ detection\\ \hline \addlinespace
        ALLQSOs &  Deep Fields  + GAMA  &    4436      &  2469  (56\%)  \\
                &  Deep Fields    &    2105      &   1555 (74\%)\\
                    &  GAMA     &    2331   &    914 (39\%)\\
        MWQSOs  &  Deep Fields  + GAMA   &  1789       &  1002 (56\%)           \\
                &  Deep Fields        &     489       & 451  (94\% )   \\
                &  GAMA      &      1300      &  551  (42\% )  \\

    \end{tabular}
    \caption{Description of the different samples used for this study, including the number of LOFAR detected quasars and their corresponding fractions. The MWQSOs and the ALLQSOs samples correspond to the areas in Figure \ref{fig:fields} which are filled with grey stars and the light-cyan shaded areas, respectively. }
    \label{tab:samples}
\end{table}

\subsection{Radio data} \label{sec:deepradio}

The main radio data used are part of the LoTSS Deep Fields DR1 \citep{tasse21, sabater21}, which comprise repeated observations of the LOFAR High Band Antenna (HBA) at 150 MHz of the ELAIS-N1, Lockman Hole, and Boötes fields.
The LoTSS Deep Fields observations are among the deepest wide-field radio-continuum survey at low-frequencies to date, at a resolution of $\sim 6$ arcsec and with an rms sensitivity reaching 20, 22, and 32 $\mu$Jy beam$^{-1}$ in the centres of the three fields, respectively (Figure \ref{fig:fields}).
To complement the LOFAR Deep Fields and obtain a larger sampled volume which is key given the low number density of quasars,  we use LOFAR equatorial observations from pipeline verification data which targets three fields from the Galaxy And Mass Assembly survey (GAMA).
Each GAMA field consists of 4 LOFAR pointings at 2x4h integration time, at a resolution of $\sim 8$ arcsec and achieving an rms noise level of around 200 $\mu$Jy beam$^{-1}$ in the centres of each field (Figure \ref{fig:fields}).  
The diversity of sensitivity limits in the different fields is overcome by both field-independent and combined assessments, as well as by focusing on a comparison strategy, carefully matching red and control samples for intrinsic AGN luminosity ($\Lsixum$) and redshift  following our previous work, \citep[Lz-matching: ][\citetalias{CR21}]{klindt19, fawcett20, rosario20}. 
In particular, the Lz-matching was performed by selecting the closest control QSOs in $\Lsixum$ and $z$ for each red quasar (within each field), matched using the Nearest Neighbour method defining a two-dimensional metric on the  $\Lsixum-z$ space. 
Whenever limited sample sizes do not allow us to apply this matching approach, we make an explicit note in the respective sections.

For the catalogued sources, the radio to optical cross-matching was performed using the likelihood-ratio matching for all fields, and additional visual inspection for the LoTSS Deep Fields. A detailed description of the  cross-matching process is presented by \citet{kondapally20}.
For this study, we use optical position coordinates from the LoTSS catalogue to find the counterparts to the SDSS QSOs in our sample, adopting a search radius of 1 arcsec with respect to the SDSS positions.
In addition to using the LOFAR catalogues associated with the data, we compute the radio fluxes for the SDSS QSOs with fainter radio emission, which are not formally detected by the source extraction strategy used in the construction of the catalogue (Python Blob Detector and Source Finder, see \citealt{pybdfs15}), which is limited to sources with SNR$\gtrsim 5$.
We estimate these fluxes by measuring the peak fluxes within circular apertures around the known SDSS source positions, and consider a measurement as a detection if it has a SNR$>3$, where the noise is estimated from the local rms.
Given the higher spatial resolution, for the Deep Fields we choose an aperture of  $r_{DF}= 6''$, while for the GAMA fields we use $r_{G} = 10''$ (chosen due to the non-circular shape of the beam in the equatorial fields).
To test the robustness of our method, we compare the fluxes computed using this approach with those from the source extraction method for the sources formally detected in the catalogue.
We find a tight agreement between these values.

\subsection{Detection fractions, radio loudness and morphology}
\begin{figure}
    \centering
    \includegraphics[ trim={0.2cm 0.2cm  2cm 0.1cm},clip, width=\linewidth]{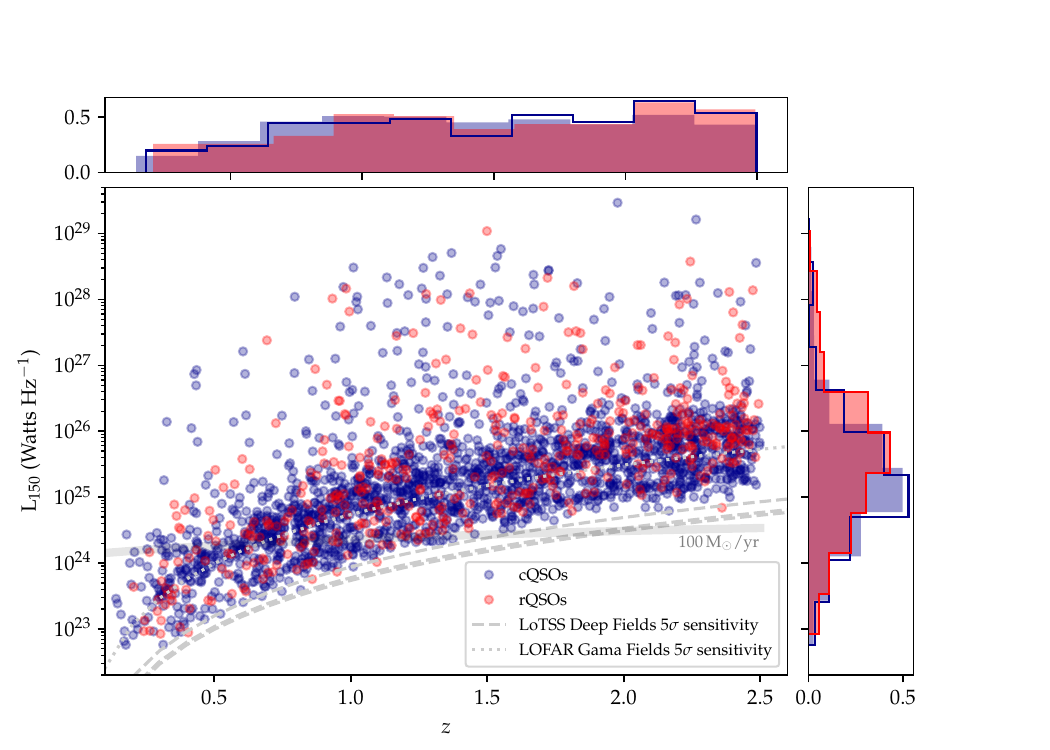}
    \caption{Luminosities at 150 MHz as a function of redshift for SDSS QSOs detected at 3$\sigma$ by LOFAR (ALLQSOs sample).  LOFAR detections in the total SDSS sample are represented as orange and skyblue markers, for red QSOs and control QSOs, respectively. LOFAR detections in the Lz-matched sub-samples are represented as red and dark blue markers for red QSOs and control QSOs, respectively. The grey shaded area represents the radio luminosity expected from a SFR of 100 M$_{\odot}$ year$^{-1}$. }
    \label{fig:LOFARlumz}
\end{figure}

From the 4436 QSOs in the ALLQSOs sample, we find that  941 QSOs (21 \%) have radio detections in the LOFAR catalogues ($\gtrsim5\sigma$, LoTSS Deep + GAMA), and 2469 (56\%) have radio detections at $> 3\sigma$.
Remarkably, in the areas of LoTSS Deep (northern areas) with multiwavelength coverage (MWQSOs), 94\%  (451/480) of the SDSS QSOs are 3$\sigma$-detected, which is the largest radio detection fraction ever achieved for optical QSOs.
Using the catalogued sources only ($\gtrsim5\sigma$), we find a detection fraction of 60\% (286/480), instead.
For the entire LOFAR region of LoTSS Deep (ALLQSOs) this number decreases to 74\%, due to the fact that the radio sensitivity decreases moving from the centers of the LOFAR pointings.
See Table \ref{tab:samples} for details on the sample differences.
The unique high detection fractions in the LoTSS Deep fields and the complementary wide field coverage of the GAMA fields, allows this study to achieve a near-complete characterisation of the radio properties of optically selected QSOs at both the radio-faint and the radio-bright ends.


In order to explore the radio luminosity distribution of the QSOs, in Figure \ref{fig:LOFARlumz} we show the 150 MHz luminosities \Llofar{}, of all radio-detected sources as a function of redshift.
The upper and right panel show histograms of the distributions of redshift and radio luminosities for the radio-detected red and control sample (shaded histograms), as well as the Lz-matched control sample (dark blue unfilled histograms; see Section \ref{sec:selection} for a description of the Lz-matched sample).
While it is clear that an improved similarity in redshift distribution (upper panel) is achieved for the red and control samples matched in redshift and 6 $\mu$m luminosity as expected, this is not the case for the radio luminosity (right panel), where the red quasar populations clearly shows a higher radio luminosity despite being matched in 6 $\mu$m luminosities.
As a reference we plot as well the sensitivity limits for each one of the LOFAR fields.
Figure \ref{fig:LOFARlumz} shows that the low-frequency radio luminosities covered by our samples of red and blue QSOs vary by 6 orders of magnitude, from $\rm 10^{23}-10^{29}\,W\,Hz^{-1}$.

\begin{figure}
    \centering
    \includegraphics[trim={ 0cm 0.4cm 2cm 1.2cm },clip, width=\linewidth]{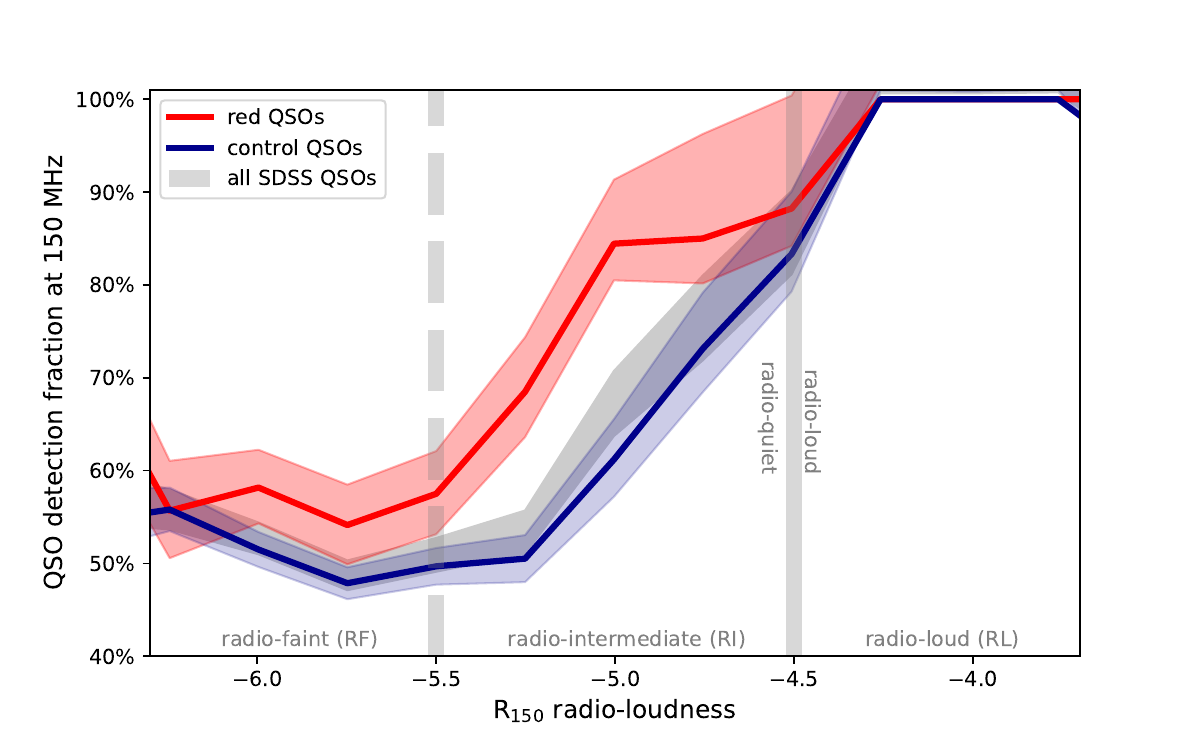}
        
    \caption{Radio detection fraction at $3\sigma$ of the ALLQSOs sample in the combined LOFAR fields, as a function of radio loudness. Samples are matched in $\Lsixum$ and $z$. Radio loudness was estimated for the sources without detections using measured radio fluxes at the QSO position. These upper limits are included in this plots and dominate the lower bins of the radio faint regime.  Consistent with previous studies, the detection fraction of red QSOs (red line), is significantly higher than the overall detection fraction (grey) and the control sample (blue) in particular in the regime of intermediate radio loudness ($-5.5<R_{150}<-4.5$). The shaded areas represent 1$\sigma$ binomial uncertainties for each bin following \citet{cameron11}.}
    \label{fig:LOFARdetect_R150}

\end{figure}

The range of radio luminosities and radio properties in the QSO population is in part driven by the diversity in AGN intrinsic power, traced by $\Lsixum$. 
Analysing and comparing radio luminosities alone can thus be non-informative on the physical conditions which lead to the production of radio emission itself.
To focus on this instead,  
we weight the radio luminosities by the AGN radiative power and define the radio loudness of the populations as
$R_{150} = 1.5 \times L_{150}[{\rm W\,Hz^{-1}}] / \Lsixum [{\rm ergs\,s^{-1}}]. $   
This definition follows the description by \citet{rosario20}, and is equivalent to the relation used at higher radio frequencies \citep{klindt19}. 
Note that in contrast to many definitions of radio-loudness in the literature which use the optical regimes to trace the radiative power in AGN \citep[e.g., ][ Arnaudova et al., submitted]{kellerman89, gurkan19}, we use L$_{6\mu \rm m}$ instead to avoid biases due to optical reddening, crucial for our study.
The deep and wide-field radio maps used in this work allow us to explore the radio loudness parameter space in a complete manner.
We calculate upper-limits for the radio-loudness in non-detected sources using radio aperture fluxes. These are included in Figure \ref{fig:LOFARdetect_R150} to showcase the detection fractions as a function of radio-loudness, where they contribute to the lower radio-loudness bins. However, we note that for the remaining investigation we only consider sources that are radio-detected. 
To put the radio-loudness distribution observed with our deep radio data in context with traditional radio classifications, we define the radio-loud/radio-quiet limit at $R_{150}=-4.5$, equivalent to previous studies.
While our sample comprise an extensive range of radio-loudness ($-7.4<R_{150}<-2.$), the vast majority of the QSOs (94\%) in our sample would be considered `radio-quiet' according to traditional classification, and only a small percentage would be considered radio-loud (6\%). Given the continuous distribution of radio loudness which is observed in our sample and other deep radio surveys \citep[see also ][Yue et al., submitted]{gurkan19, macfarlane21}, for the remaining of this paper we move on from the traditional bimodal scenario to a more detailed study of the radio-loudness distribution.

To investigate how the radio detection fraction changes as a function of radio loudness in our QSO sample, Figure \ref{fig:LOFARdetect_R150} shows the radio detection fraction of red QSOs and the control sample populating each bin of radio loudness.
In agreement with our previous studies \citep{klindt19, fawcett20, rosario20}, we find a clear difference in the detection fractions as a function of radio loudness $R_{150}$ for red and control QSOs, where red QSOs achieve close to 85\% detections already at $(R_{150}>-5.2)$, whereas control QSOs only reach that detection fraction at the radio-loud limit $(R_{150}>-4.5)$.
In the regime traditionally known as `radio-quiet' $(R_{150}<-4.5)$, the radio detection of red QSOs is higher than for the control sample, with a maximum difference at a radio-loudness of $(R_{150}\sim -5.0)$.
To further investigate the radio emission in QSOs traditionally classified as `radio-quiet' \citep[$R_{150}<-4.5$; e.g., ][]{klindt19, rosario20}, we further divide the radio-quiet population into two regimes which appear relevant from Figure \ref{fig:LOFARdetect_R150}.
We therefore define the radio-intermediate regime within the range of $-5.5<R_{150}<-4.5$, and the radio-faint regime at values of $R_{150}<-5.5$.
Indeed, we observe that the radio-detection excess for red QSOs peaks in the radio-intermediate regime and then decreases for the radio-faint sources.
This suggests that the radio-loudness parameter might be a good indicator of the different mechanisms producing radio emission. 
Moreover, sources within the radio-intermediate regime appear to be particularly relevant for the study of the connection between QSO reddening and radio emission. 

\begin{figure}
    \centering
        \includegraphics[width=0.8\linewidth]{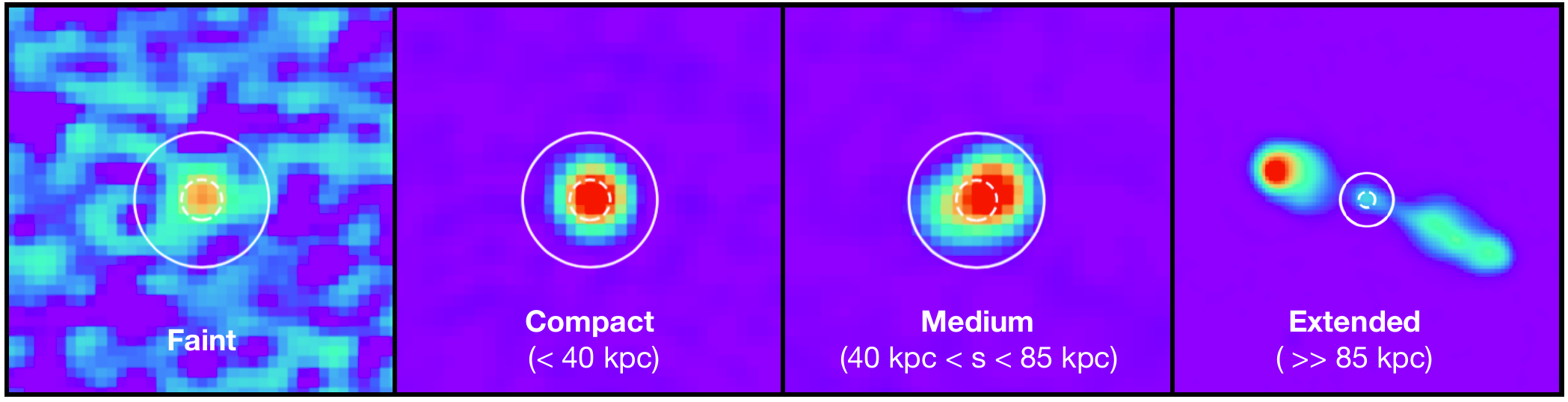}
        \includegraphics[width=\linewidth]{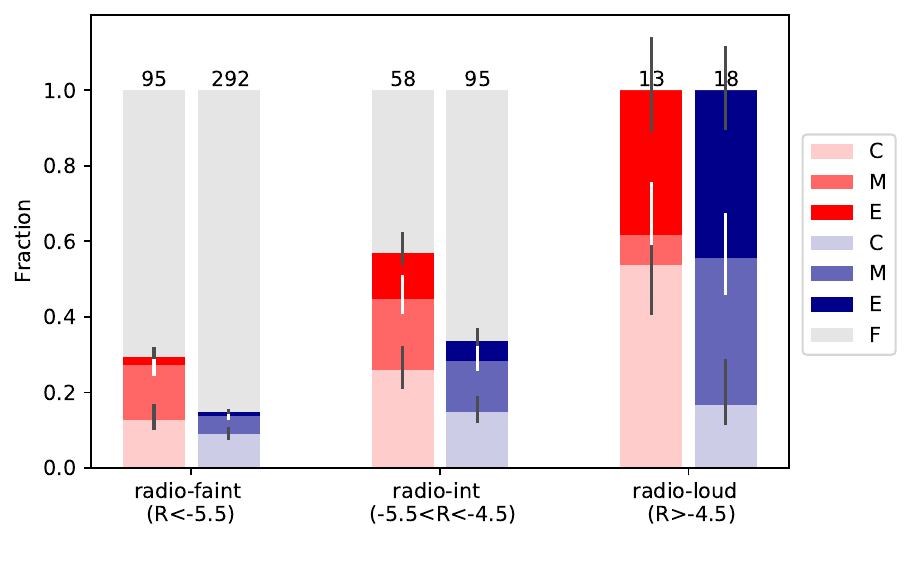}
    \caption{ Fraction of sources of a given radio morphology (faint, compact, medium and extended) within each bin of radio loudness. 
     Thumbnails of the LOFAR images are shown in the upper panel to exemplify the morphological classes used here, where two circles are depicted at 3'' (dashed line) and 10'' (solid line).
   The bars in the lower panel include all morphology classes (each denoted by a single letter).
    Only ALLQSOs from the LoTSS Deep Fields catalogue (northern fields) were used here due to their depth and better resolution (6''). Error bars represent 1$\sigma$ binomial uncertainties.}
    
    \label{fig:morph}

\end{figure}

The morphology of radio sources can be informative on the physical origin of the radio emission. 
To apply a morphological analysis to our sample, we restrict our coverage to the QSOs with counterparts in the LoTSS Deep Fields catalogues (e.g., $\gtrsim 5\sigma$; 166 red QSOs and 405 control QSOs) due to their superior depth and spatial resolution ($6''$).
Extracting thumbnail images from the LOFAR maps at the QSO positions, we classify the sources  through visual inspection.
To make a simple comparison we classify the sources into four morphological classes.
These are defined as: `compact'  sizes, which are unresolved at the scales of the LoTSS Deep resolution, equivalent to scales of $< 50$ kpc for a redshift of $z=2$; `medium' sizes, which are resolved, slightly larger than the LOFAR beam and corresponding to sizes of  40 kpc $< s <$85 kpc; `extended' sizes which are long multi-component structures, corresponding to $>$85 kpc; and `faint', as shown in the upper panel of Figure \ref{fig:morph}.
The respective numbers and fractions are reported in Table \ref{tab:morphs}.
In particular, the latter class `faint' refers to sources with low surface brightness and low SNR, where no assessment can be made on the morphology.
As shown in the lower panel of Figure \ref{fig:morph}, we find that the majority of QSOs are too faint (55\% and 77\% for red and control QSOs) to characterise their morphologies.
After excluding the faint sources, we find that QSOs are overall preferably compact (46 \%).
This is in agreement with our previous studies \citep{klindt19, fawcett20, rosario20} using high frequency and/or shallower radio data.
The fraction of sources with medium sizes is lower (35\% and 37\% for the red and control QSOs, respectively), whereas a minority are extended (19\% and 17\% for the red and control QSOs, respectively). 
Given the unique radio sensitivity achieved in the LoTSS Deep Fields, we can now investigate the distributions of the morphologies as a function of radio loudness. 
Given that the large majority of morphologies in the radio-faint end is predominantly unconstrained, we focus on the other two radio-loudness bins.
Overall we find that compact and medium (marginally resolved) morphologies are present for both red and control populations in a similar manner in the radio-intermediate regime, suggesting that radio mission at both sub-galactic and circum-galactic scales can be both expected.
In both the radio-intermediate and at the radio-loud end, we find that compact morphologies are more prevalent for red QSOs than for the control QSOs.
This is particularly clear in the radio-loud bin where $\sim 60\%$ of the red QSOs remain compact, while only $\lesssim 20\%$ of the control sample does so, suggesting that red QSOs are less likely to host radio structures of 10s of kpcs.

\begin{table}[]
    \centering
    \begin{tabular}{c|c c|c c}
          &  \multicolumn{2}{c|}{red QSO}   & \multicolumn{2}{c}{control QSO}  \\
          \hline \addlinespace
    All &  393  &    &  1712 &   \\    
    Radio-det &  166  &  (42.2\%)  &  405   &  (23.6\%)   \\   
    Faint &  92  & (23.4\% )  &  313   &   (18.3 \%)  \\
    Compact & 34   & (8.6 \%)    &43    & ( 2.5\%)   \\ 
    Medium &   26 &   (6.6 \%)  & 33    &  (1.9\%) \\
    Extended &  14  &  (3.5\%) & 16    &   (0.9\%)  \\

    \end{tabular}
    \caption{Number (and fraction) of QSOs included in each of the morphology classes in Figure \ref{fig:morph}. The QSO sample used here is restricted to the ALLQSOs sample within the LoTSS Deep Fields (northern fields), and only radio counterparts from the catalogue ($\gtrsim 5\sigma$) were considered due to the high SNR requirements of the morphological study. }
    \label{tab:morphs}
\end{table}

Our morphology results are in overall agreement with the conclusions from our previous work, i.e. we also find that faint and compact sources dominate the morphology distribution of QSOs \citep{klindt19, fawcett20, rosario20, rosario21}, and that the radio-excess in red QSOs is mainly driven by the compact and marginally resolved sources \citep[][]{fawcett20}.
However, our deep low-frequency radio data reveal larger fractions of medium-sized (marginally resolved) sources ($35\%$ and $37\%$ for red and control QSOs, respectively) as compared to the fractions reported by \citet{fawcett20} using similarly deep radio data at $1.4-3$ GHz ($18\%$ and $23\%$ for red and control QSOs, respectively).
These different fractions are potentially explained by the fact that the most extended regions of radio structures are more visible at lower radio frequencies, as these have steeper radio spectra due to ageing of the electron populations producing the radio emission.
We note however that this analysis is limited by caveats related to the small sample statistics as well as potential biases intrinsic to the visual inspection which are not taken into account within the uncertainties.

\section{Results}

\subsection {The origin of the bulk of the radio emission in QSOs is not star formation}\label{subsec:SFR}

\begin{figure}
    \centering
    \includegraphics[ width=\linewidth]{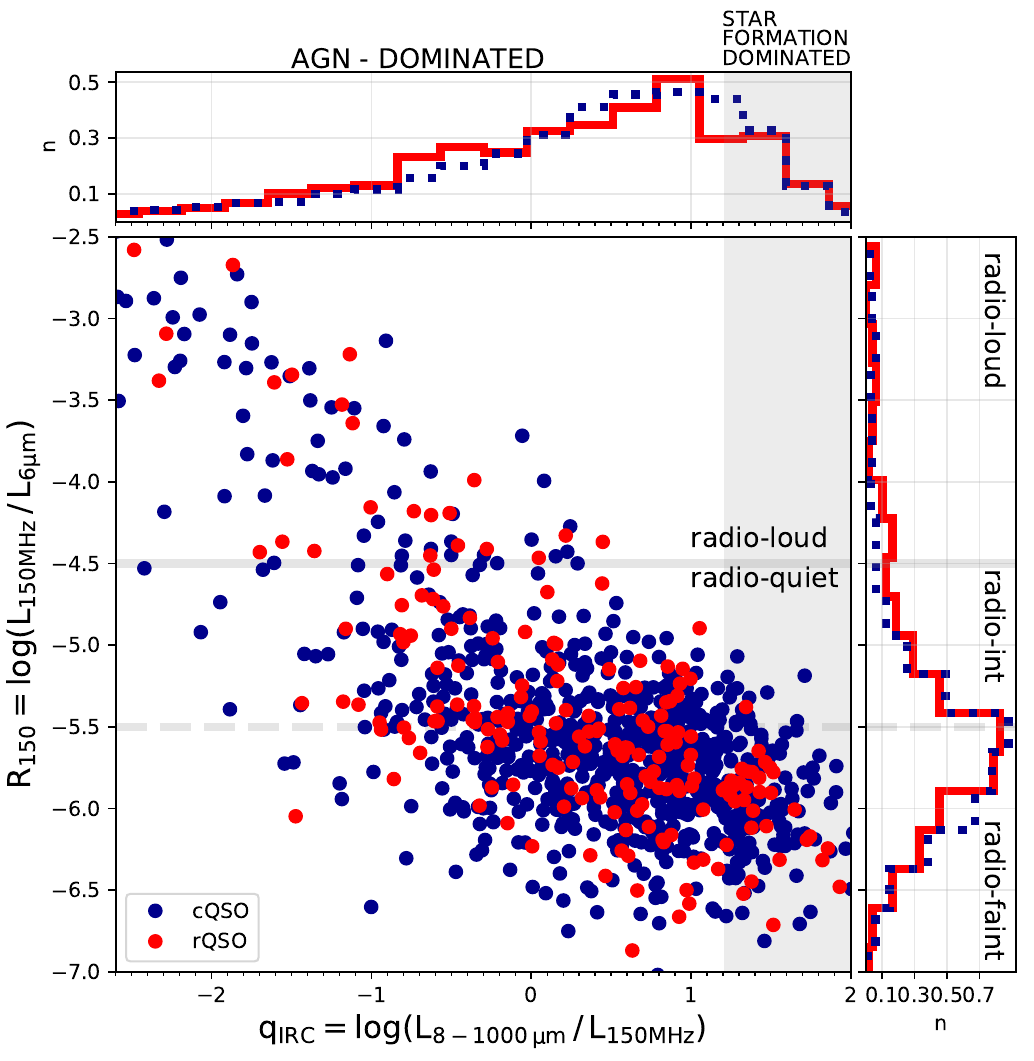}
    \caption{The radio loudness R$_{150}$ as a function of the q$_{\rm IRC}$-value, which parametrises the IR-radio correlation, for the MWQSO sample. The central panel shows the posterior median values for red QSOs and the control sample. The upper panel shows the combined distribution of the q$_{\rm IRC}$ posteriors (100 realizations per source) for the two samples, obtained from the Bayesian SED fitting. The right panel shows the combined distribution of the R$_{150}$ values for the two samples.}
    \label{fig:IRC}
    \vspace{-0.5cm}

\end{figure}

It has been widely debated in the community whether synchrotron emission from star formation can be the principal or a significant source of radio emission in radio-quiet AGNs \citep[see] [for a recent review]{panessa19}.
Building upon previous work \citep{fawcett20, rosario20}, we now take advantage of the unique sensitivity and sky coverage of our radio data, to explore this question in a statistical sample of QSOs that populates the whole range of radio luminosities.
Based on our detailed Bayesian SED-fitting approach applied to deep photometry in all the studied fields, including deep \textit{Herschel} PACS and SPIRE data, we can estimate the radio synchrotron emission expected from star formation using the infrared-radio correlation (IRC).
Since the SED fitting output is required for the central and upper panel of Figure \ref{fig:IRC},  we use here the MWQSO sample, which has extensive deep photometric coverage allowing us to distinguish between the different host galaxy and AGN components in individual QSOs.

The IRC is a tight empirical correlation observed across several orders of magnitudes in radio luminosities, redshift, and using different radio frequencies \citep[for low radio frequencies see e.g.,][]{cr17, gurkan18,  smith21, mccheyne22, best23}. 
The IRC is parametrized by the  q$_{\rm IRC}$ parameter, defined as q$_{\rm IRC}= \log(({\rm L}_{8-1000 \mu{\rm m}}/ 3.75\times 10^{12} \rm Hz)/{\rm L}_{150{\rm{MHz}}})$, where L$_{8-1000 \mu{\rm m}}$ is the galaxy rest-frame IR luminosity (in erg s$^{-1}$) and is estimated after subtracting the contribution from the AGN torus emission to the total infrared emission.
L$_{150 \rm MHz}$ is given in ergs s$^{-1}$ Hz$^{-1}$, and the factor of 3.75$\times 10^{12}$ Hz is the frequency corresponding to 80 $\mu$m used in the definition to make q$_{\rm IRC}$ a dimensionless quantity.
The AGN subtraction to the total IR luminosity is estimated from the Bayesian SED fitting and, while it varies largely across the sample, the torus contribution is overall substantial with a median value of the posterior distribution of $\sim 70\%$, and is equivalent for both red and control QSOs (see \citetalias{CR21} for more details).

The central panel in Figure \ref{fig:IRC} shows the median posterior values of the q$_{\rm IRC}$ parameter for each source in relation to their radio loudness R$_{\rm 150}$.
To estimate the fraction of the radio emission that is consistent with star formation, we use the value of q$_{\rm IRC}$ at 150 MHz found by \citet{cr17} for star forming galaxies  q$_{\rm IRC-150 MHz}=1.54$  with a scatter of 0.5, shown as a grey shaded area in Figure \ref{fig:IRC}.
The upper side panel shows the distribution of q$_{\rm IRC}$ for red and control QSOs. 
In particular, we note that the q$_{\rm IRC}$ distribution in the side panel is a superposition of the posterior distributions for each source, obtained from the Bayesian SED fitting, therefore accounting for the uncertainties in the measured IR luminosities, as well as uncertainties on the AGN subtraction.
To compare the distributions in the different samples we apply the non-parametric two-sample Kolmogorov–Smirnov (KS) test.
Based on this test, no significant differences are seen in the relative contribution of star formation to the radio emission between red and control QSOs ($p_{\rm KS}$-value= 0.48), albeit the sample size might limit the recognition of subtle differences. 
This is in line with \citetalias{CR21}, which report no differences in the star formation rates between the red and the control QSOs.
The right  side panel in Figure \ref{fig:IRC} shows the distribution of R$_{\rm 150}$ for the MWQSO sample. 
Using the MWQSO sample shown in Figure \ref{fig:IRC}, no significant differences are found between the  R$_{\rm 150}$ distributions for red and control QSOs according to the KS test, due to the small samples involved.
In contrast, using the ($>2$x) larger ALLQSO sample, we find significant differences between the  R$_{\rm 150}$ distributions for red and control QSOs ($p=0.002$ overall, $p=10^{-7}$ when Lz-matched).
Using this larger sample it is clear that red QSOs have a higher density at intermediate radio-loudness, and lower density at the radio-loud end, as already indicated by the detection fractions in Figure \ref{fig:LOFARdetect_R150}.
As seen in Figure \ref{fig:IRC} for the MWQSO,  a trend between q$_{\rm IRC}$ and R$_{\rm 150}$ is observed, where higher radio-loudness in the QSOs makes it less probable for their radio emission to originate in star formation.
At the radio-faint end (R$_{\rm 150}<-5.5$), star formation can start dominating  the radio emission for a small fraction of the QSOs.

\begin{figure}
    \centering
    \includegraphics[trim={ 0.5cm 0.3cm 2cm 0.5cm},clip, width=\linewidth]{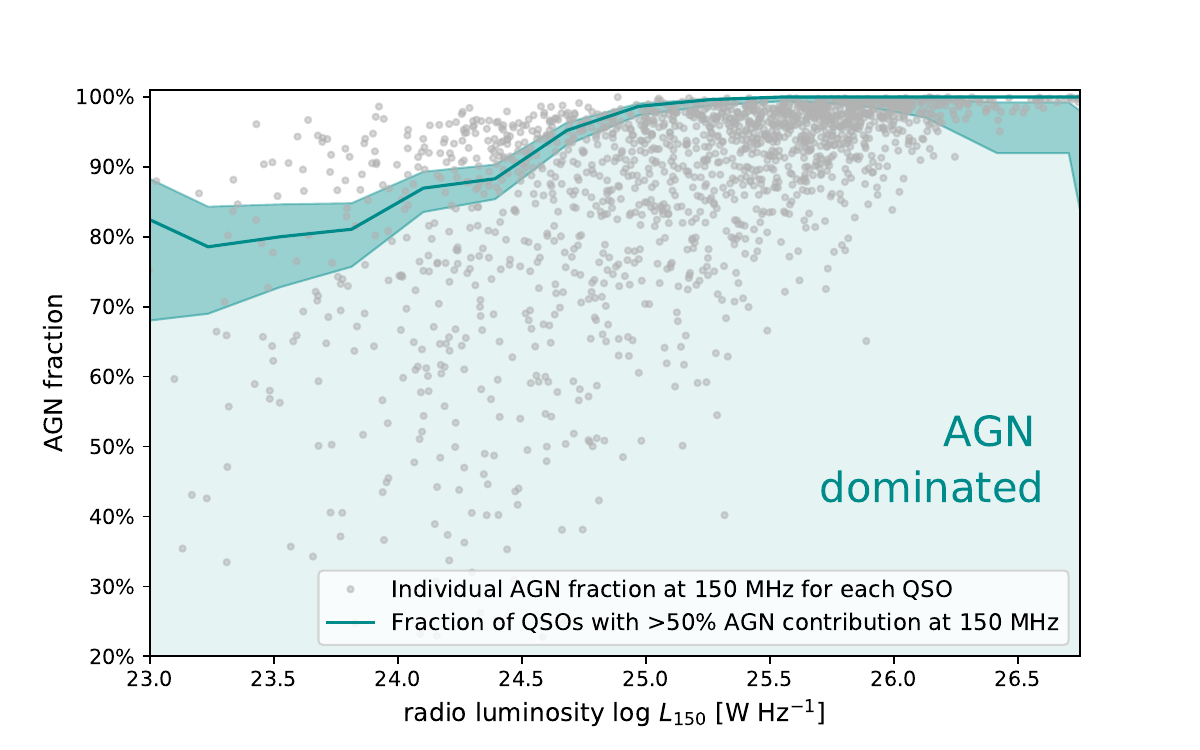}
    \caption{The fraction of AGN contribution to the radio emission of QSOs at 150 MHz, as a function of radio luminosity L$_{150}$ (for the MWQSO sample). The grey dots show the AGN fraction of the L$_{150}$ for each source, i.e. the fraction which is in excess from the SF relation accounting for the scatter. The dark cyan line and shaded area represent the percentage of the overall population which have AGN fractions larger than 50\%, i.e., are dominated by the AGN. The shaded dark area around the line shows 1-$\sigma$ binning uncertainties.}
    \label{fig:AGNfrac}
    \vspace{-0.5cm}
\end{figure}

To further investigate the origin of the radio emission in QSOs, we use the q$_{\rm IRC}$ to compute the fraction of the radio luminosity of each source which originates in the AGN, i.e. which is in excess from the IRC.
Since a multiwavelength characterisation is required here, the study in this section is restricted to the MWQSOs sample.
As is apparent from Figure \ref{fig:IRC} only a small fraction of the QSO population has radio emission which is consistent with being produced by star formation (grey shaded area).
We discuss this fraction in detail next, as well as the implications of different calibrations of the q$_{\rm IRC}$ value.
Assuming the median value of  q$_{\rm IRC}$ reported by \citet{cr17} (q$_{\rm IRC}$=1.54),  we find that the radio emission from 76\% of the QSOs in our sample ($\log \Lsixum> 43.7$) is AGN-dominated. 
However, the value and potential dependencies of q$_{\rm IRC}$ have been a matter of discussion and revision in the recent years enabled by the increasing number of deep radio surveys.
Indeed, \citet{cr17} reported a redshift-evolution of q$_{\rm IRC}$ following the relation q$_{\rm IRC}=1.72 \times (1+z)^{-0.22}$.
If we apply this relation, the fraction of QSOs where the radio emission is dominated by the AGN decreases to 70\%.
Nonetheless, larger studies using deeper radio data and mass-complete star forming galaxy samples (instead of a radio-selected sample as used by\citealt{cr17}) have reported slightly different values of q$_{\rm IRC}$ at low frequencies, and crucially, a dominant dependence on stellar mass \citep[][but see also Das et al. submitted]{smith21, mccheyne22}.
In particular, \citet{mccheyne22} reported a weaker q$_{\rm IRC} - z$ relation following q$_{\rm IRC}= 1.94 \times (1+z)^{-0.04}$ and a q$_{\rm IRC} - \rm M_{*}$ relation following q$_{\rm IRC}= 2 -0.22 \times (\log M_{*} - 10.05)$.
Applying this relation to our sample in combination with stellar masses for all sources estimated from the SED fitting (\citetalias{CR21}), we find that the fraction of QSOs where the radio emission is dominated by the AGN increases to 91\% and 93\%, for the $z$ and $\rm M_{*} - $relations, respectively.
To account for the diversity of these results, we will conservatively assume the median value reported by \citet{cr17} for the remaining of this study, i.e. that at least 76\% of all QSOs in our sample ($\log \Lsixum> 43.7$) are AGN-dominated. 

The combination of a statistical sample of QSOs, uniquely deep radio imaging and FIR-to-UV SED fitting in this study allows us to confirm and complement our previous results on SDSS QSOs \citep{fawcett20,rosario20}, which find similar conclusions albeit with smaller radio samples and less multiwavelength coverage, respectively.
To see how this fraction evolves as a function of radio luminosity, in  Figure \ref{fig:AGNfrac} we show the AGN fractions for each QSO in our sample as grey points. 
Although there is a significant variation from source to source, the bulk of the radio luminosity from the QSO population clearly originates in processes other than star formation.
A statistical view is shown with the dark cyan line and uncertainties, which represent the fraction of QSOs in our sample for which the AGN fraction at 150 MHz is higher than 50\%, i.e. the radio emission is dominated by AGN processes.
Accounting for the binning uncertainties (shown as shaded area), we find that the radio emission in QSOs originates in the AGN in $\gtrsim 68\%$ of the QSOs even at the lowest luminosity bin probed ($L_{150 \rm MHz}>10^{23}$ W Hz$^{-1}$).   
This result is also in agreement with previous literature \citep{zakamska14}, but now exploring lower AGN luminosities.
 \citet{macfarlane21} and Yue et al. (submitted) also address this question by performing a parametric modelling of the radio luminosity distributions of QSOs using larger shallower data sets from LoTSS DR2, with the model assumption that the radio emission is a superposition of two components: star formation and jet emission.
While they also find that the fraction of the radio emission in QSOs originating in jets/AGN can be significant and prevalent for QSOs at radio luminosities of $L_{150 \rm MHz}>10^{25}$, they find lower fractions at lower luminosities.
This discrepancy is potentially due to the significantly different methodologies, as well as data sets and definitions employed in these studies. In particular, their methodology consists of modelling the shape of the radio luminosity distributions while we study the emission from single sources. Second, while the data set employed in their study is significantly larger (>40000 QSOs), their radio data is shallower and the study is restricted to the radio, while our study uses deeper radio data in combination with detailed radio-to-UV photometry and multiwavelength SED modelling for each source. Finally, they define AGN dominated sources as sources which have AGN radio emission 5 times larger than that from SF, whereas we consider AGN dominated sources as those where the AGN is responsible for at least 50\% of the total radio emission.

\subsection{A fundamental connection between dust-reddening and radio emission}\label{subsec:redradio}

\begin{figure}
    \centering
    \includegraphics[trim={ 0.5cm 0.4cm 2cm 0.2cm},clip, width=\linewidth]{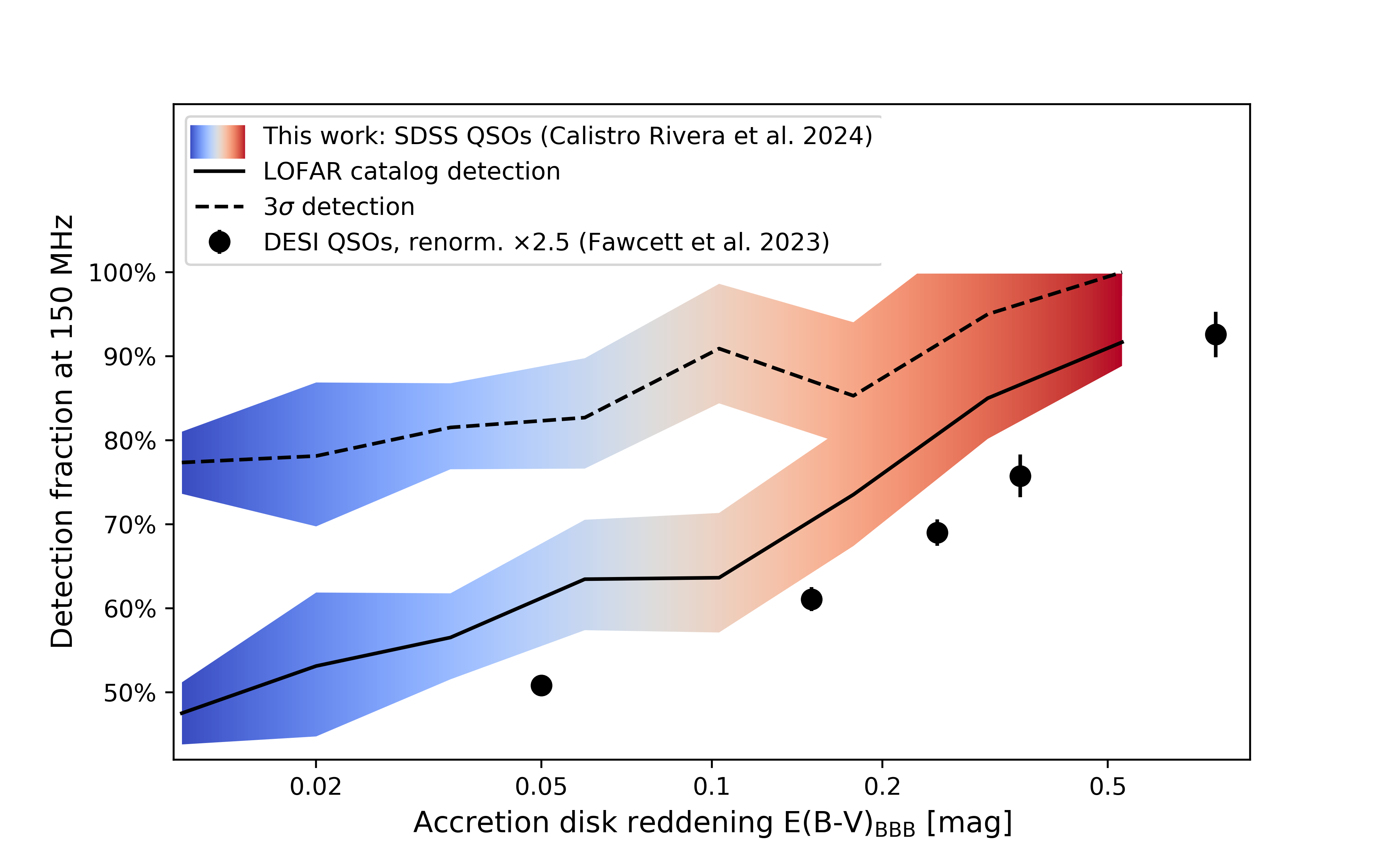}
    \caption{Radio detection fraction of QSOs at 150 MHz as a function of the accretion-disk reddening in the MWQSO sample. The accretion-disk reddening is estimated from the SED fitting for each source and is parametrised by the E(B-V)$_{BBB}$ parameter. The solid line shows the trend considering only catalogued sources  ($\gtrsim$ 5 $\sigma$), whereas the dashed line considers all QSOs detected down to $\sim$ 3$\sigma$. The shaded area represent 1$\sigma$ binomial uncertainties for each bin. A clear positive relation is apparent, where the radio detection fraction increments with increasing dust-reddening, in agreement with QSOs detected in the DESI survey (black circles).}
    \label{fig:detvsEBV}
    \vspace{-0.5cm}

\end{figure}

The remarkable finding of an enhanced compact radio detection fraction in red QSOs \citep{klindt19, fawcett20, rosario20, glikman22} suggests a connection between the radio emission and intrinsic QSO properties.
We investigate this further, by looking beyond the red and control QSO classification, and investigating the overall dependence of the radio detection fraction on dust-reddening of the accretion disk, parametrized by the E(B-V)$_{\rm BBB}$ parameter.
Through the Bayesian SED-fitting applied on the rich optical photometry available in our fields (Section \ref{sec:selection}), we have inferred robust E(B-V)$_{\rm BBB}$ for all QSOs in our sample (\citetalias{CR21}). 
Interestingly, Figure \ref{fig:detvsEBV} shows a positive correlation between the detection fraction as a function of reddening E(B-V)$_{\rm BBB}$. The solid black line shows the dependence for the  sources in the catalogue (5$\sigma$), while the dashed black line, which shows higher fractions for the 3$\sigma$ detections at 150 MHz, as expected, presents a similar trend. 
Uncertainties from the binning are shown as shaded area, colour coded to represent the continuous QSO colour transition traced by the E(B-V)$_{\rm BBB}$.
As shown in Figure \ref{fig:detvsEBV}, this result is in agreement with recent findings by \citet{fawcett23}, estimated from fitting the DESI spectra of $\sim$35000 QSOs. 
Similar to our findings, they report a continuous relation between the radio detection fraction and the E(B-V) values out to higher dust extinction values than we can probe in SDSS QSOs (up to E(B-V)$\sim$ 1.0) but with lower-sensitivity LOFAR data.
The finding of this continuous relation strongly suggests that there is a physical relation between the dust attenuation at nuclear scales (as suggested by the high dust temperatures found by \citetalias{CR21}) and the production of radio emission, which is not restricted to the most reddened optical sources but is a continuous property relevant for the entire quasar population. 

\subsection{The connection between radio, dust reddening and outflows } \label{subsec:radiodustout}

In our previous work (\citetalias{CR21}), our analysis of SDSS spectral properties revealed a higher incidence of winds in red QSOs as compared to the control sample, traced by high-velocity [\textsc{Oiii}] wing components ($z<1$), and \textsc{Civ} blue-shifts ($1.4<z<2.5$).
Additionally, we found that the sources with stronger outflow components exhibit higher levels of hot nuclear dust emission ($\rm T_{\rm dust} \sim 1000 \, K$), revealed by the SED fitting as MIR excess emission which was not recovered by dusty torus models. 
These high dust temperatures provided indirect evidence that the dust reddening in QSOs originates over nuclear and circumnuclear scales ($<$ 1 kpc) and is connected to \textsc{Civ} and [\textsc{Oiii}] winds, therefore suggesting the reddening material might be distributed along nuclear dusty winds, similar to what has been observed in a few local AGN \citep[e.g.,][]{honig17, lyu18}.
To find a link between the higher incidence of winds in red QSOs with their higher radio detection fractions, we now explore the radio data in parallel using the ALLQSOs sample.

Studies in the literature suggest that the radio detection fraction in QSOs increases as a function of wind velocities  \citep[e.g., \textsc{Civ} blueshifts, ][]{rankine21}, or that radio detected AGN have a higher outflow rate than non-detected radio AGN \citep[][Escott et al., in prep.; Petley et al., submitted]{mullaney13, perrotta19}, although most of them do not consider colour in their study.
We evaluate this in our sample using catalogue-detections ($\sim$5$\sigma$) in the ALLQSOs samples, and indeed we find that the radio detection fractions grow as a function of of both increasing \textsc{Civ} blueshifts and increasing [\textsc{Oiii}] velocities. 
However, as also discussed by \citet{rankine21}, we argue that this effect is largely dominated by a dependence of both radio detections and \textsc{Civ} and [\textsc{Oiii}] line-widths on AGN bolometric luminosities \citep[see also][]{woo17}.
To test this and capitalize on the depth of our data, we use the radio-loudness parameters instead of radio detection fractions, which normalizes the radio luminosities by the AGN intrinsic radiative power.

Building upon our analysis in \citetalias{CR21} for these same sources, Figure \ref{fig:outflow_radio} shows the cumulative distributions of spectral line velocities used as outflow tracers ([\textsc{Oiii}] wing velocities and \textsc{Civ} blue-shifts), for both red and control QSOs.
We note that to obtain these spectral properties in \citet{rakshit20}, a double Gaussian model was used to represent the [O~{\sc iii}] emission line; one for the core narrow component and another for the wing component, to characterise potential winds. 
In this paper we use the widths of the wing component to characterise the wind velocities.
In the case of C~{\sc iv} the total profile was used without the subtraction of a narrow component, because of the ambiguity in the presence of narrow components in these lines.
A detailed description of the spectral fitting can be found in \citet{rakshit20}.
We apply the same sample cuts as in our previous work, to  avoid contamination from the host galaxy  and  incompleteness ($\Lsixum$>44.5 \ergs ), as well as to ensure high quality detections of the wing components ($\rm F_{[\textsc{Oiii}]peak}>3\times rms_{5100\AA}$).
After these cuts we have in total 33 red QSOs and 144 control QSOs for the [\textsc{Oiii}] outflow study and 108 red QSOs and 921 control QSOs for the \textsc{Civ} outflow study.
Given the small sample size for the [\textsc{Oiii}] study, we are not able to match the red and control sources by $\Lsixum$ and $z$, however we note that the cuts applied already ensure very similar distributions.

We now split our sample into three sub-samples of different levels of radio-loudness, radio faint ($R<-5.5$), radio-intermediate ($-5.5<R<-4.5$) and radio-loud ($R>-4.5$).
We overplot the cumulative distributions of non-detected sources just for reference to make the comparison easier.
As shown in the upper panel of Figure \ref{fig:outflow_radio}, we find a slight apparent enhancement in [\textsc{Oiii}] wind strength for the red QSOs in the radio-faint regime as compared to the control sample, yet it is not significant according to the 2-sample KS test (p-values $= 0.16$).
Interestingly, for the radio-intermediate subsample, we find that the [\textsc{Oiii}] wind strength for the red QSOs significantly increases as compared to the control sample (p-value $= 0.02$).
For the last bin of radio-loud sources, our sample was severely reduced, therefore not allowing us to make any comparison.
These findings suggests that the bulk of increased [\textsc{Oiii}] wind velocities reported in the red QSOs by \citetalias{CR21}, is tightly connected to radio emission in red QSOs with intermediate radio-loudness.

In the lower panel of Figure \ref{fig:outflow_radio}  we perform the same experiment for the \textsc{Civ} blue-shifts.
We find increased \textsc{Civ} blue-shifts for the red QSOs as compared to the control QSO, also when matching the samples in  $\Lsixum$ and $z$, with p-values of $p_{\rm ks}= 0.0001, 0.002,0.035, 0.001$ for the radio non-detected, the radio-faint, the radio-intermediate and the radio-loud populations, respectively. 
This is in line with our previous findings in \citetalias{CR21}, as well as with simulations which propose that radiation pressure on dust is effective in launching outflows \citep[e.g., ][]{costa18, soliman23}.
Interestingly, in contrast to what we found in the [\textsc{Oiii}] analysis, the \textsc{Civ} velocity distributions do not seem to change with respect to the radio-loudness. 
Indeed, comparing the distributions in the three radio-loudness bins to the distributions for the radio-undetected sources (dotted lines), we find that they are consistent according to the KS-test ($p_{ks}=0.98, 0.66 ; p_{ks}=0.99, 0.24 ; p_{ks}=0.56, 0.11$ for control and red QSOs at the radio-faint, the radio-intermediate and radio-loud bins, respectively).
This suggests that while there might be an increase in radio detections with increased \textsc{Civ} blue-shifts \citep[e.g.][]{rankine21}, this effect disappears when accounting for the intrinsic AGN power.
We will discuss possible scenarios which could explain the differences in the outflow properties between \textsc{Civ} and [\textsc{Oiii}] in Section \ref{sec:discussion}.

We note, however, that our results are limited by a few factors.
The SDSS [\textsc{Oiii}] spectra are available only for sources at $z<1$, and strictly require further SNR cuts due to the limited spectral quality, which significantly decreases the sample relevant to the [\textsc{Oiii}] studies.
While the compared quasar samples have equivalent median intrinsic luminosities (differences are $<0.1$ dex) and similar redshift distributions due to the cuts performed, the limited sample size did not allow us to further match them in $z$ and $\Lsixum$.

\section{Discussion and Conclusions} \label{sec:discussion}

\begin{figure}
    \centering
    \includegraphics[trim={ 0cm 0.3cm 0cm 0cm},clip, width=\linewidth]{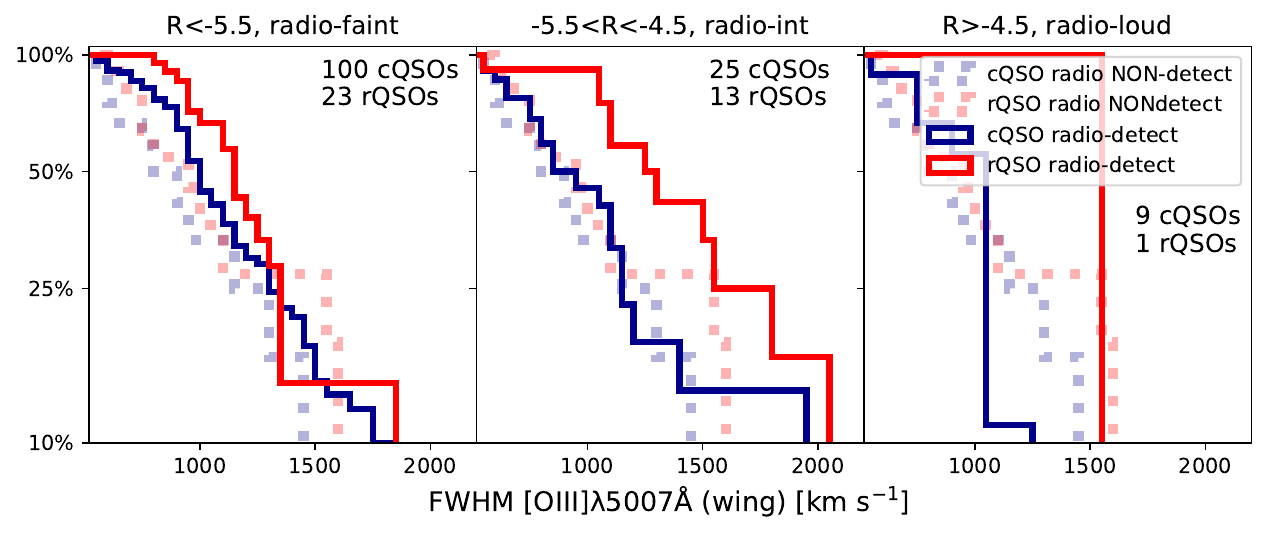}
    \includegraphics[trim={ 0cm 0.3cm 0cm 0cm},clip, width=\linewidth]{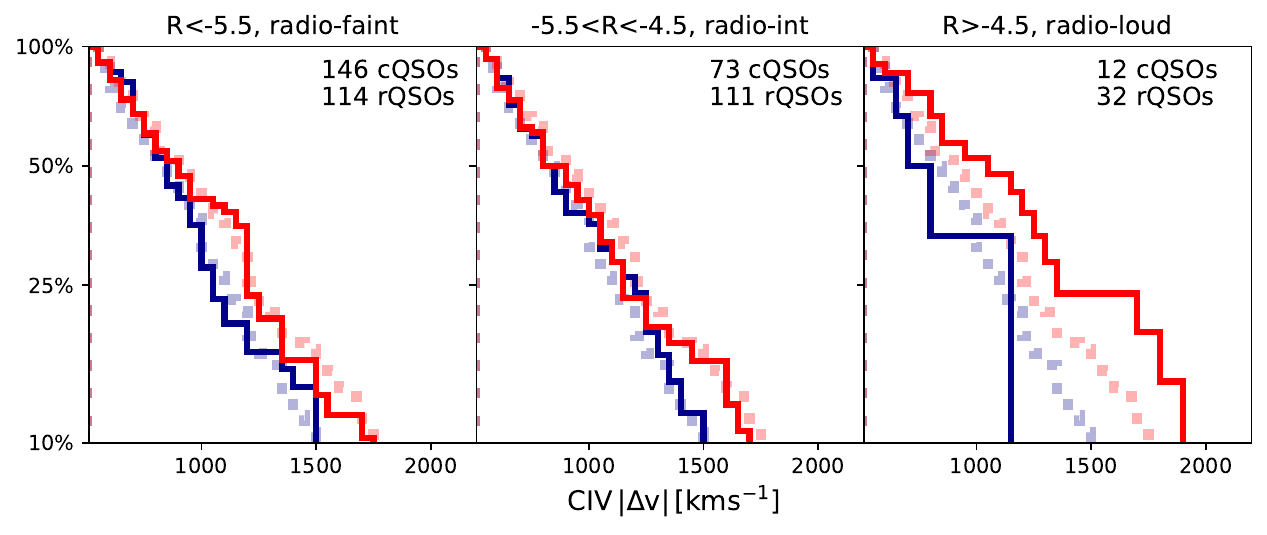}
    \caption{Cumulative distributions of the [\textsc{Oiii}] and \textsc{Civ} velocities (upper and lower panel, respectively). The [\textsc{Oiii}] velocities are parametrised by the FWHMs for the wing component of the [\textsc{Oiii}] emission line, and the  \textsc{Civ} velocities are paremetrised by the velocity blueshifts of the \textsc{Civ} emission lines. Red and blue lines represent the distributions for the red and control QSOs, respectively. The left to right panels show the distributions for radio-faint, radio-intermediate and radio-loud subsamples. Solid lines are the distributions of radio detected QSOs within each bin of radio-loudness, while the distributions of undetected sources are shown as dotted lines for reference.
    Here we use the ALLQSOs sample at the redshifts required to cover the respective emission line within the SDSS spectral window ($z<1$ for the [\textsc{Oiii}] line, and $1.4<z<2.5$ for \textsc{Civ}). Additionally, a luminosity threshold from our parent sample was applied to avoid contamination from the host galaxy, as well as a SNR criterion to ensure high quality detections of the [\textsc{Oiii}] wing components.
    }
    \label{fig:outflow_radio}

\end{figure}

In this paper we have presented a comprehensive study of the origin of the radio emission in SDSS QSOs, capitalizing on radio surveys of unprecedented sensitivity and sky coverage, as well as a consistent multiwavelength characterisation through SED modelling and spectral fitting.
Our unique radio data set, in combination with the extensive multiwavelength study presented by \citet{CR21} for the same sources, allow us to consistently investigate the physical connection between radio emission, dust reddening and the presence of outflows in a statistical sample of QSOs. 
Our main results are the following:
\begin{itemize}
    \item Using the deepest radio maps to date, available for a subset of our sources, we find detection fractions up to 94\% showing that the radio emission from QSOs is virtually ubiquitous. 
    \item Applying AGN-tailored IR-to-UV SED-fitting to isolate the host-galaxy and AGN contributions to the overall energetics, we find that the radio emission in $>76\%$ of QSOs is consistent with originating from AGN processes rather than star formation at galactic scales.
    \item Consistent with previous studies, we find that this AGN radio emission is enhanced for red QSOs as compared to the control sample, in particular at intermediate radio-loudness and at compact scales, unresolved by the resolution limit of our survey ($<6''$). 
    \item Going beyond the color-based classification, we find a continuous positive correlation between the radio detection fraction and the accretion disk reddening parameter E(B-V)$_{\rm BBB}$ in QSOs. This finding suggests that the production of radio emission is potentially physically connected to the dust attenuation at circumnuclear scales, and is an intrinsic property in QSOs.
    \item We find that the [\textsc{Oiii}] and \textsc{Civ} outflow velocities generally increase when QSOs have red colours (\citetalias{CR21}). We find that the [\textsc{Oiii}] wind velocities are significantly larger for red QSOs at intermediate radio-loudness, whereas for the control QSO sample no differences are observed with respect to radio loudness. Similarly, no differences are observed for the \textsc{Civ} emission with respect to radio properties.
\end{itemize}

We can now contextualise these observations within physical scenarios that potentially provide a unified explanation for all these phenomena. The consistent relationship between reddening and the radio detection fraction, as presented in Figure \ref{fig:detvsEBV}, implies the existence of a shared physical mechanism responsible for both the compact radio emission in QSOs and the reddening of the accretion disk emission.
Having excluded star formation as the origin of the compact radio emission through our SED-fitting approach in Section \ref{subsec:SFR}, we evaluate the remaining potential explanations.

A possible candidate for this mechanism is shocks from dusty radiation-driven winds \citep[e.g., ][]{costa18, soliman23}. These AGN winds have been proposed in the literature  to explain the radio emission in radio-quiet quasars \citep{zakamska16, nims15, hwang18}. 
In particular, accelerated particles in the shock fronts of the winds emit synchrotron radiation through the interaction with the magnetic fields.
Since no physical explanation is apparent which would enhance the radio emission in shocks for dusty (as compared to less dusty) winds of fixed wind velocities, we would expect stronger wind signatures for stronger radio emission, irrespective of their colour. 
However, for red QSO  we detect a significant enhancement of \textsc{[Oiii]} wind velocities at intermediate radio-loudness, as compared to the radio-faint regime,  whereas no enhancement is observed for the control QSO sample (Figure \ref{fig:outflow_radio}).
This suggest that circumnuclear dust is a key condition to detect a relation between radio emission and \textsc{[Oiii]} outflows.
Moreover, our morphology analysis in Figure \ref{fig:morph} suggests that,  at the intermediate radio-loudness regime, moderately extended emission (>40 kpc) are present in the morphologies of the radio emission in both red and control QSOs.
While we cannot rule out the wind scenario for red QSOs given their primarily compact morphologies, we argue that the radio emission  at 40-85 kpc scales observed for a fraction of both the control and red (radio-quiet) QSOs cannot be due to shock interactions (which would be expected to be most effective on host-galaxy scales where the gas density is higher) and is almost certainly due to jets.
Therefore, to reconcile this observation with the observed continuity in radio-loudness, as well as the continuous radio-reddening relation found in Figure \ref{fig:detvsEBV} and by \citet{fawcett23}, a self-consistent explanation is achieved by invoking the presence of compact radio jets of different sizes in QSOs.
The ubiquitousness of radio jets of compact and moderate scales in QSOs is indeed supported by recent direct observations of such compact radio jets at high-resolution \citep[e.g., ][]{an12, jarvis19, hartley19} as well as by modelling the flux density distributions in deep wide-field radio surveys \citep[e.g., ][Yue et al., submitted]{macfarlane21}.

The extended morphologies (>40 kpc) observed in a fraction of the QSOs in the radio-intermediate and radio-loud bins (Figure \ref{fig:morph}) strongly suggest that these structures are radio jets, since those extensions cannot easily be created by winds if they are not collimated or exceptionally strong, and since we find no strong wind signature in the \textsc{[Oiii]} spectra for the control QSOs in these radio intermediate regimes (Figure \ref{fig:outflow_radio}). 
Interestingly, red QSOs at the radio-loud regime are still dominated by compact emission (<40 kpc). 
Indeed, we argue that the connection between the radio emission, reddening, and outflows occurs predominantly on circumnuclear scales, based on the following observations.
First, the results presented in \citetalias{CR21} find dust temperatures of 1000K for the dust responsible for the reddening in this QSO sample. These temperatures suggest that the dust is distributed on scales of a few pc to a few 10s of pc.
Additionally, the compact radio emission in red QSOs appears to potentially  dominate at scales of $<2$ kpc, as suggested by the results reported in the literature at higher radio resolution at 1.4 GHz \citep{rosario21}
Finally, we find that the dust and the radio emission is connected to the presence of high-velocity \textsc{[Oiii]} winds, i.e. wind components at circumnuclear narrow line region (NLR) scales, whereas we do not find this connection to the $\textsc{Civ}$ winds prevalent at broad line region (BLR) scales. This therefore suggests that the radio-wind connection occurs at circumnuclear scales rather than happening at the accretion disk/jet base scales. 

This suggests that (a) the origin of the radio emission in red QSOs is different to the control QSOs, and/or (b) radio jets in red QSOs are in a younger phase (i.e., they are smaller) and/or (c) the radio jets in red QSOs are maintained compact due to interaction with the circumnuclear/ISM environment.
While scenario (a) is plausible, it does not agree well with the continuous relation which we observe between the radio emission and QSO reddening in Figure \ref{fig:detvsEBV}, which requires a unified physical explanation for the radio emission in red and control QSOs. So, while different mechanisms might indeed be in place, scenarios that include radio jets as a continuous physical component are more likely, even if these are potentially not the most relevant for the radio excess observed to increase with reddening.

Scenario (b), where red QSOs host younger (more compact) radio jets than control QSOs, on the other hand,  would explain our observations on morphology well, as well as those found in the literature \citep{rosario21}, since sizes of radio jets are a proxy of the age of the jet \citep[e.g., ][]{hardcastle20}.
It would also be in agreement with other studies on red QSOs, which suggest red QSOs are part of a different evolutionary phase \citep{urrutia05, glikman12, glikman13, banerji15, klindt19, glikman22}.
In order to explain our observations on dust and outflows, on the other hand, the younger red QSOs would be required to be undergoing a blow-out phase at the same time, where dust and gas would be ejected from the obscured AGN phase to an unobscured phase, as suggested by evolutionary models \citep{sander88, hopkins08, alexander12}. 
This is supported by our findings and studies in the literature, which also find that the [\textsc{Oiii}] outflows are more prevalent when the radio emission is compact \citep{molyneux19}, and young, as suggested by their peaked radio SEDs \citep{santoro20, kukreti23}.
Our observations of a continuous relation between dust reddening and radio detection, however, is more suggestive of a causal and gradual connection than the parallel occurrence of jet launching and a blow-out phase.
We explore this in scenario (c).

Finally, in scenario (c) we propose that the radio emission arise from a superposition of compact radio jets and their interaction with the environment, which produce additional radio emission from jet-induced shocks.
For this scenario, we consider that red QSOs are sources with overall denser and dustier circumnuclear environments than the control QSOs, potentially due to evolutionary reasons connected to (b).
The interplay between compact radio jets and this dense circumnuclear environment is naturally expected to produce outflows and shocks \citep[e.g. in simulations,][]{mukherjee18, bicknell18}.
The increased incidence of jet-induced outflows and shocks in redder QSOs would, in turn, produce both the radio emission observed to gradually increase as a function of E(B-V)$_{\rm BBB}$ (Figure \ref{fig:detvsEBV}), as well as the stronger \textsc{[Oiii]} kinematics we observe in these radio-intermediate sources \citep[Figure \ref{fig:outflow_radio} and e.g., ][]{mizumoto23}.
Additionally, these interactions with a dense environment would restrict the growth of the evolving  radio jet, therefore maintaining it to the compact scales at which they are observed even at the radio-loud regime \citep[here $<40$ kpc due to resolution, while sub-arsec resolution studies suggest sizes of $< 2$ kpc are typical at least at 1.4 GHz][]{rosario21}.
This confinement to compact scales will also enhance the radio luminosity of the jets by decreasing adiabatic expansion losses  \citep[e.g., ][]{barthel96}.
Conversely, the radio-jets in the control QSOs, residing in less dense environments, can potentially pierce through a circumnuclear and interstellar medium without significant interaction with the galactic environment \citep[as seen also in simulations][]{mukherjee18}, growing to the moderate sizes observed in the radio-intermediate and radio-loud populations.
While extended radio jets are observed in a similar manner for the control QSOs and red QSOs, they are overall very rare, and potentially correspond to a small fraction of sources which are sufficiently powerful to pierce quickly through the (potentially clumpy) surrounding medium despite the denser environments. 
This would mean that the early evolution of  compact radio jets in QSOs would either be truncated by the dense medium and remain compact (such as most red QSOs even in the radio-loud regime), or potentially grow without much interaction but disappear as it evolves within/shortly after surpassing galactic scales (such as the control QSOs), while a small minority would evolve into a `jetted', radio-loud object.
Scenario (c) can be also discussed in the evolutionary context described in (b).
In red QSOs, the proposed compact jets could expel dust and gas from the nuclear regions, reducing the dust and gas column density in the process of growth, till they become bluer, consistent with a blow-out phase.

While we cannot rule out other scenarios, our results are consistent with a picture where small-scale jets are a ubiquitous phase in QSOs \citep[see also ][]{jarvis19, hartley19, macfarlane21}, and that they represent a key part of the evolutionary path of QSOs. 
Moreover, radio emission from shocks produced by the interaction of these small-scale jets with denser environments can potentially explain the observed continuous relation between reddening and radio detection fractions in quasars. 
If the ubiquitousness of compact radio jets is confirmed, this would represent  a crucial aspect to consider in our understanding of quasar feedback in galaxy evolution \citep[e.g., ][]{thomas21}. 
Indeed, small-scale jets are suggested to have a larger impact on galaxy growth than their large-scale counterparts, heating and inducing turbulence in the surrounding ISM more efficiently, and therefore suppressing the star formation capabilities of the ISM. \citep[e.g.,][]{mukherjee18, cielo18, girdhar22}. 
Further studies are necessary to confirm this model and fully understand  the connection between radio emission, dust reddening and outflows in QSOs.
In particular, larger studies of spectral properties will be insightful to better characterise outflows in statistical QSO samples.
Ongoing programs, such as WEAVE-LOFAR \citep{smith16} will provide large new samples of radio-detected QSOs in the near future.
Radio SED curvature and spectral analysis (Fawcett et al in prep., Sargent et al in prep.), as well as high-resolution sub-arsecond imaging of these sources (e.g. LOFAR LBA, VLBI observations) will furthermore inform us on the origin of the radio emission by estimating the time scales of the emission and resolving the scales at which it is produced.


\section{ACKNOWLEDGEMENTS}
For the purpose of open access, the author has applied a Creative Commons Attribution (CC BY) licence to any Author Accepted Manuscript version arising from this submission.
GCR acknowledges the ESO Fellowship Program and a Gruber Foundation Fellowship grant sponsored by the Gruber Foundation and the International Astronomical Union. 
DMA thanks the Science Technology Facilities Council (STFC) for support from the Durham consolidated grant (ST/T000244/1).
PNB, RK and BY are grateful for support from the UK STFC via grant ST/V000594/1.
DJBS acknowledges support from the UK Science and Technology Facilities Council (STFC) under grant ST/V000624/1.
MIA acknowledges support from the UK Science and Technology Facilities Council (STFC) studentship under the grant ST/C005460/1.
LKM is grateful for support from UKRI [MR/T042842/1].
IP acknowledges support from INAF under the Large Grant 2022 funding scheme (project “MeerKAT and LOFAR Team up: a Unique Radio Window on Galaxy/AGN co-Evolution”).

\bibliographystyle{aa}
\bibliography{LOFAR_QSO.bib}

\end{document}